\documentclass[iop]{emulateapj}

\usepackage[squaren, Gray, cdot]{SIunits}
\usepackage{float}
\usepackage{multirow}
\usepackage{stmaryrd}
\usepackage{scalerel}
\usepackage{mathtools}
\usepackage{color}

\def\stretchint#1{\vcenter{\hbox{\stretchto[220]{\displaystyle\int}{#1}}}}
\def\bs{\!\!}

\def\mearth{M_\oplus}
\def\rearth{R_\oplus}

\slugcomment{Accepted for publication in The Astrophysical Journal}

\begin{document}

\title{Constraints on Super-Earths Interiors from Stellar Abundances}

\author{
B. Brugger\altaffilmark{1},
O. Mousis\altaffilmark{1},
M. Deleuil\altaffilmark{1},
and F. Deschamps\altaffilmark{2}
}

\altaffiltext{1}{Aix Marseille Univ, CNRS, LAM, Laboratoire d'Astrophysique de Marseille, Marseille, France  (bastien.brugger@lam.fr)}
\altaffiltext{2}{Institute of Earth Sciences, Academia Sinica, 128 Academia Road Sec. 2, Nangang, Taipei 11529, Taiwan}

\begin{abstract}

Modeling the interior of exoplanets is essential to go further than the conclusions provided by mean density measurements. In addition to the still limited precision on the planets' fundamental parameters, models are limited by the existence of degeneracies on their compositions. Here we present a model of internal structure dedicated to the study of solid planets up to $\sim$10 Earth masses, i.e. Super-Earths. When the measurement is available, the assumption that the bulk Fe/Si ratio of a planet is similar to that of its host star allows us to significantly reduce the existing degeneracy and more precisely constrain the planet's composition. Based on our model, we provide an update of the mass-radius relationships used to provide a first estimate of a planet's composition from density measurements. Our model is also applied to the cases of two well-known exoplanets, CoRoT-7b and Kepler-10b, using their recently updated parameters. The core mass fractions of CoRoT-7b and Kepler-10b are found to lie within the 10--37\% and 10--33\% ranges, respectively, allowing both planets to be compatible with an Earth-like composition. We also extend the recent study of Proxima Centauri b, and show that its radius may reach 1.94~$\rearth$ in the case of a 5~$\mearth$ planet, as there is a 96.7\% probability that the real mass of Proxima Centauri b is below this value.

\end{abstract}

\keywords{Earth --- planets and satellites: composition --- planets and satellites: individual (CoRoT-7b, Kepler-10b, Proxima Centauri b) --- planets and satellites: interiors}

\section{Introduction}
\label{sec:1}

The huge diversity of discovered worlds, in terms of physical and orbital parameters, led to enlarge the planet population with new families. Among them, Super-Earths and sub-Neptune bodies fill the gap between the terrestrial planets and the giant gaseous planets that compose our solar system. Measurements of the mass and radius of an exoplanet, mostly obtained from radial velocity and transit methods, respectively, allow to derive the body's mean density. This quantity gives a rough estimate of the planet's bulk composition, whether the mean density is closer to that of the Earth (5.51 g/cm$^3$) or to that a gaseous planet like Jupiter (1.33 g/cm$^3$). To better constrain the composition of exoplanets, that is the distribution of elements inside these bodies, interior models have been developed, based on our knowledge of the properties of the Earth and other solar system bodies \citep{Valencia06,Sotin07,Seager07,ZengSeager08,Rogers10,Dorn15}. Models of planetary interiors inherently present a degeneracy issue as two planetary bodies with different compositions may have the same mass and radius. For instance, the same set of mass and radius allows interior models to generate both a planet displaying properties similar to those of the Earth (silicate mantle surrounding a relatively small metal core) and a planet possessing a larger core with a smaller mantle surrounded by a thick water layer.

In this paper, we present an interior model derived from those developed for the Earth, and able to handle compositions as various as those of small planets and large satellites of the solar system (Mercury-like to ocean worlds). This model is limited to the case of dense solid planets with possible addition of water, and does not consider planets that harbour thick gaseous atmospheres made of H/He. Using a more appropriate equation of state compared to previous studies, we provide up-to-date mass-radius relationships for planets with masses under 20~$\mearth$. Coupled to an adapted numerical scheme, our model explores the parameter space for the possible compositions of solid planets to minimize the set of compositions consistent with the measured physical properties. We aim at breaking the aforementioned degeneracy by incorporating the Fe/Si bulk ratio of the investigated planet into our code. This ratio helps constraining the size of the metal core inside the planet, a parameter mostly concerned by this degeneracy.

The numerical model is described in Section~\ref{sec:2}. Section~\ref{sec:3} presents our results concerning the investigation of the interiors of two well-known low-mass exoplanets, namely CoRoT-7b and Kepler-10b, based on the latest estimates of their physical parameters, and assuming they do not contain gaseous envelopes. We also investigate the interior of Proxima Centauri b by considering a larger mass range than the minimum mass considered in a previous study \citep{Brugger16}. Section~\ref{sec:4} is dedicated to discussion and conclusions.

\section{Model}
\label{sec:2}

\subsection{Planet internal structure}
\label{ssec:2.1}

\begin{figure}
\begin{center}
	\includegraphics[width=0.9\columnwidth]{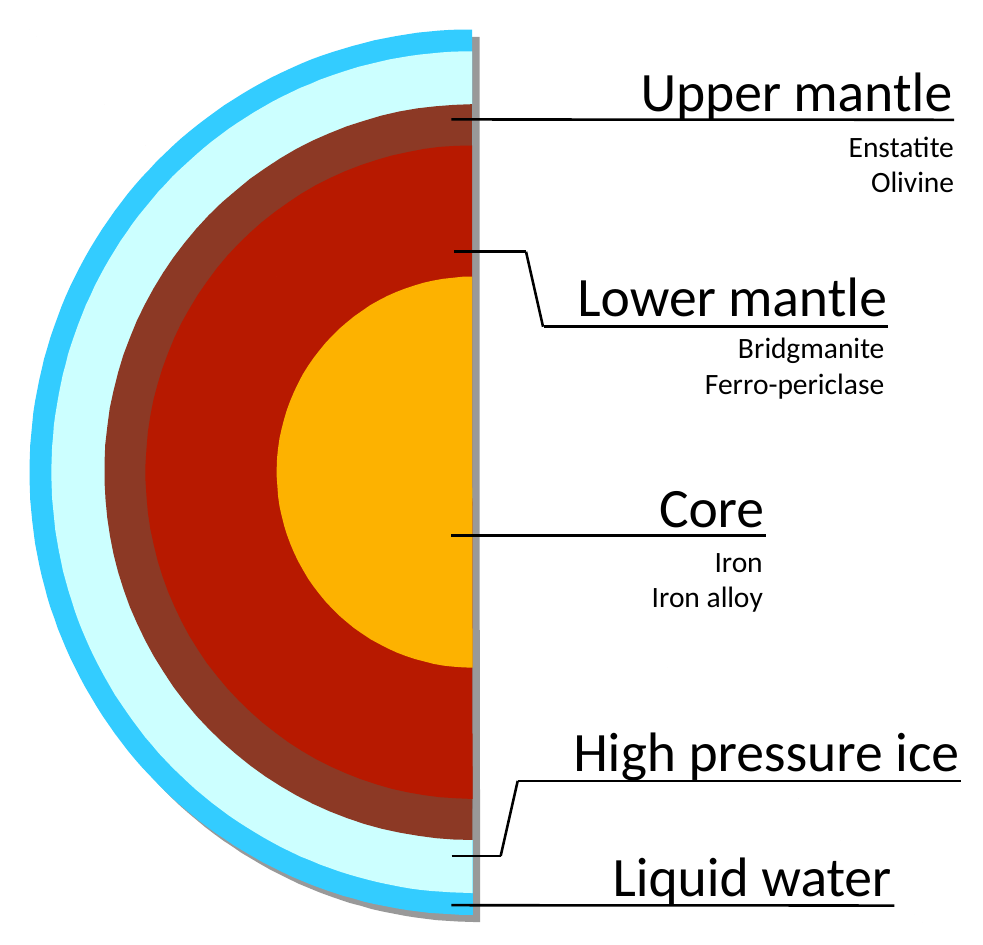}
	\caption{Schematic view of the different concentric layers that compose our interior model: metallic core, lower and upper silicate mantles, high-pressure water ice, and liquid water. Depending on the mass of the water layers, the upper mantle or the high pressure ice layer may be absent.}
	\label{fig:1}
\end{center}
\end{figure}

Our model is based on the approach described by \cite{Sotin07}. It assumes a fully differentiated planet with several shells (or layers). Reflecting the terrestrial planets in our solar system, the three main layers consist of a metallic core, a silicate mantle, and a hydrosphere. In our model, the mantle and the hydrosphere can be divided into two sublayers each, leading to planets that can be made of up to a total of five concentric layers (see Figure~\ref{fig:1}), which are from the center:

\begin{enumerate}
	\item \textbf{The core.} In the Earth, the core is essentially composed of iron, along with smaller fractions of other metals (as nickel or sulfur). Here we assume that it consists in a single layer formed of a mixture of pure iron (Fe) and iron alloy (FeS).
	\item \textbf{The lower mantle.} Here, the elements Fe, Mg, Si and O form the silicate rocks. The lower mantle corresponds to a region of high pressure \citep{DziewonskiAnderson81} in the phase diagram of silicates, which can be in the forms of bridgmanite (Mg,Fe)SiO$_{3}$ and ferro-periclase (Mg,Fe)O (also known as perovskite and magnesiow\"ustite, respectively).
	\item \textbf{The upper mantle.} This layer is made of the same elements as in the lower mantle, but in the form of olivine (Mg,Fe)$_{2}$SiO$_{4}$ and ortho-pyroxene enstatite (Mg,Fe)$_{2}$Si$_{2}$O$_{6}$ because of a lower pressure \citep{DziewonskiAnderson81}. 
	\item \textbf{The high-pressure water ice.} As for the mantle, the hydrosphere divides into two layers: because some planets may include a significant amount of water, a layer of water ice VII can form at pressures reaching several GPa \citep{Frank04}. Note that if the ice VII layer is thick enough the pressure at its bottom may be too large for olivine and enstatite to exist, in which case the upper mantle would be absent.
	\item \textbf{The liquid water.} On top of ice VII, water is in liquid form, provided that the surface conditions of the planet are close to the Earth's values.
\end{enumerate}

These five layers, represented in Figure~\ref{fig:1}, allow to model the interiors of planets with various compositions, from terrestrial (i.e. fully rocky) planets like the Earth or Mercury, to ocean planets that possess a massive amount of water (like the icy moons of the jovian and saturnian systems). The differences between two planetary compositions are determined by the masses and sizes of these layers. In practice, the model's necessary inputs correspond to the masses of the three main layers (core, mantle, and hydrosphere), since their distribution into the distinct sublayers can be computed from the phase change laws of the different materials. These inputs can be expressed in terms of $x_{\rm core}$, $x_{\rm mantle}$ and $x_{\rm water}$, corresponding to fractions of the planet's total mass for the core, mantle and hydrosphere, respectively. From mass conservation, we get $x_{\rm mantle} = 1 - x_{\rm core} - x_{\rm water}$, allowing one to derive the planet's interior from the {\it a priori} knowledge of its total mass and the values of the core mass fraction (CMF) $x_{\rm core}$ and water mass fraction (WMF) $x_{\rm water}$.

Three additional parameters set the distribution of all chemical species in the different layers: the fraction of alloy in the core $f_{\rm alloy}$, the overall Mg/Si ratio of the planet $\left(\frac{\rm Mg}{\rm Si} \right)_{\rm P}$, and the amount of iron present in the silicate mantles $\rm Mg\# \equiv \left( \frac{\rm Mg}{\rm Mg+Fe} \right) _{\rm Mantle}$, describing the level of differentiation of the planet \citep{Sotin07}. In the absence of compositional data for the host stars of investigated planets, we use those derived from the Earth (see Table~\ref{tab:1}).

\begin{deluxetable}{lcl}
	\tablecaption{Compositional parameters and surface conditions for the Earth \citep{Stacey05,Sotin07}}
	\tablehead{
		\colhead{Parameter} & 									\colhead{Value} 	& \colhead{Description}
	}
	\startdata
		$x_{\rm core}$											& 0.325				& Core mass fraction						\\
		$x_{\rm water}$											& 0.0005			& Water mass fraction						\\
		$f_{\rm alloy}$											& 13\%				& Fraction of alloy in the core				\\
		$\left( \frac{\rm Mg}{\rm Si} \right) _{\rm P}$			& 1.131				& Corrected Mg/Si ratio						\\
		$\left( \frac{\rm Fe}{\rm Si} \right) _{\rm P}$			& 0.986				& Corrected Fe/Si ratio						\\
		$\rm Mg\#$												& 0.9				& Mg number									\\
		$T_{\rm surf}$ (K)										& 288				& Surface temperature						\\
		$P_{\rm surf}$ (bar)									& 1					& Surface pressure							\\
	\enddata
	\label{tab:1}
\end{deluxetable}

Once the composition and the mass of the planet are set, an iterative process solves the canonical equations for gravitational acceleration $g$, pressure $P$, temperature $T$, and density $\rho$, until convergence is reached, and provides a planet radius. This process, as well as the convergence conditions, are detailed in the Appendix.

\subsection{Equations of state}
\label{ssec:2.2}

An equation of state (EOS) gives the dependence of density to pressure and temperature. It is specific to each material through its thermodynamic and elastic parameters. Most EOS are obtained by fitting the measurements of $\rho$ versus $P$ and $T$ in laboratory experiments. This fit is then extrapolated outside the measurement range to reach pressure values existing inside the Earth and more massive planets.

\cite{Sotin07} used the well-known Birch-Murnaghan EOS, in its third-order development (see Appendix), where the thermal dependency of the pressure $P$ is directly incorporated in the coefficients. The third-order Birch-Murnaghan EOS (hereafter BM3 EOS) is often used because it is particularly fast for computations. However this EOS is limited to low pressure values ($P <$~150--300~GPa; \citealp{Seager07}) and to temperatures of the same orders of magnitude as those encountered inside the Earth's mantle \citep{Sotin07}. Above these values, the results from this EOS may deviate from theory \citep{Valencia09}. Therefore, \cite{Sotin07} used the BM3 EOS in the upper mantle and the liquid water layer, where the pressure remains low. In the deeper layers, the authors used the Mie-Gr\"uneisen-Debye (MGD) formulation as a replacement for BM3, which better describes the behavior of the used materials. The MGD EOS dissociates static pressure and thermal pressure, by adding a thermal Debye correction to the isothermal Mie-Gr\"uneisen formulation (see Appendix).

In our model, we replace the MGD EOS by another EOS already used by \cite{Valencia07a}, namely the Vinet EOS (see Appendix; \citealp{Vinet89}). As for the MGD EOS, the Vinet EOS separates static and thermal pressures with the use of the thermal Debye correction. By doing so, the Vinet EOS presents the same advantages as the MGD formulation in comparison to the BM3 EOS. However, although these three EOS have the same validity range in pressure, the Vinet EOS has been shown to better reproduce experimental data than other EOS, and to better extrapolate at pressures higher than $\sim$100~GPa \citep{HamaSuito96,Cohen00}. Therefore, it is well-adapted to the modeling of Super-Earths, where such high pressures can easily be reached. We however keep the BM3 formula within the hydrosphere, since the pressure and density remain relatively low in the upper two layers.

\subsection{Water model}
\label{ssec:2.4}

\begin{figure}
\begin{center}
	\includegraphics[width=\columnwidth]{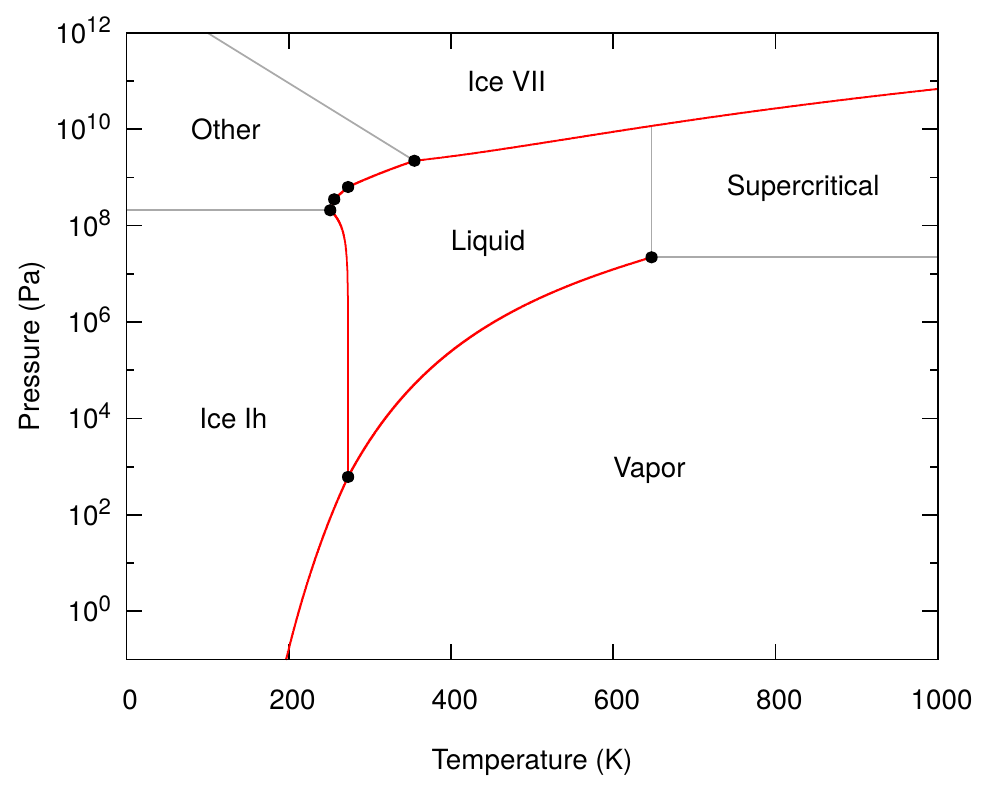}
	\caption{Simplified version of the water phase diagram used in the model to compute the boundaries between the different water layers. Black dots are triple points of water, and red curves delimit the phase transitions fitted from experimental data. Light grey lines represent the phase boundaries not considered in our work.}
	\label{fig:2}
\end{center}
\end{figure}

We consider two phases of water: ice VII and liquid water. For planets with large amounts of water, ice VII appears at pressures higher than $\sim$1 GPa. The conditions for the existence of liquid water are based on a simplified version of the phase diagram of water (see Figure~\ref{fig:2}). The liquid--ice VII transition law is taken from \cite{Frank04} who fitted a larger range of pressures (from 3 to 60~GPa) than previous works. The remaining phase change laws are taken from \cite{WagnerPruss02} and the website of the International Association for the Properties of Water and Steam (IAPWS)\footnote{\url{http://www.iapws.org/relguide/MeltSub.html}}.

\subsection{Exploration of the ternary diagram}
\label{ssec:2.5}

In our approach, the planet's mass is first set to its measured value. The compositional parameters $f_{\rm alloy}$, $\left( \frac{\rm Mg}{\rm Si} \right) _{\rm P}$, and $\rm Mg\#$ do not have a significant influence on the planet's radius compared to its mass, the CMF and the WMF \citep{Sotin07,Valencia06}. They are then fixed by default to the Earth's values. The CMF and WMF are then the remaining free parameters as they cannot be measured. They are both varied within the [0--1] range of values and linked by the relation $x_{\rm core} + x_{\rm mantle} + x_{\rm water} = 1$. The parameter space formed by the variation of these three variables is represented by a ternary diagram (see Figure~\ref{fig:4}) displaying the mass fractions of the three main layers forming an exoplanet: core, mantle, and hydrosphere (or water) \citep{Valencia07b}. Each point on the ternary diagram corresponds to a unique planet composition given by the pair (CMF,WMF). For instance, an Earth-like composition corresponds to (CMF,WMF) = (32.5\%,0.05\%) \citep{Stacey92,Stacey05}, and a Mercury-like composition to (CMF,WMF) = (68\%,0\%), even if these values are still under debate \citep{Schubert88,HarderSchubert01,Spohn01,Stacey05}. Mercury is completely dry, and the Earth's WMF is close to zero, as for all terrestrial planets in our solar system. Several moons of Jupiter and Saturn are not dry, and present a significant water amount, like Titan, with (CMF,WMF) = (0\%,50\%) \citep{Tobie06}.

\begin{figure}
\begin{center}
\includegraphics[width=0.9\columnwidth]{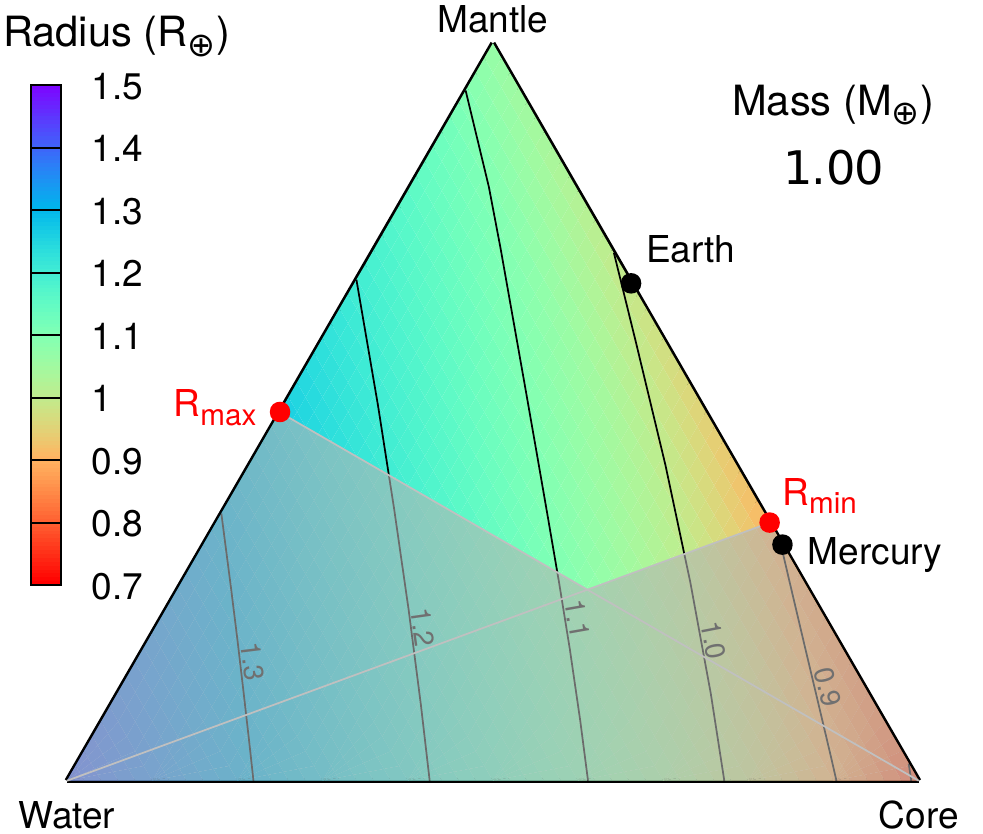}
\caption{Colormap of the planet's radius on the ternary diagram, as a function of its composition for a given mass (here $M_{\rm P} = 1 \mearth$). The computed planet radii range from 0.79~$\rearth$ to 1.40~$\rearth$. The black lines represent isoradius curves with a separation of 0.1~$\rearth$. Note that the point denoting the Earth's composition does not lay on the 1~$\rearth$ isoradius curve (see Appendix).}
\label{fig:3}
\end{center}
\end{figure}

A numerical scheme allows the model to explore the entire domain of planetary compositions formed by the ternary diagram, which produces a colormap of computed planet radii (see Figure~\ref{fig:3}). Isoradius lines drawn on this colormap illustrate the degeneracy existing in models of internal structure, as two planets may have the same mass and radius but not the same composition (i.e. they are located at different points in the diagram).

\subsection{Physical limitations on planetary compositions}
\label{ssec:2.6}

As suggested by \cite{Valencia07b}, an upper limit of 65\% can be set on the CMF, assuming that the bulk Fe/Si ratio of the considered planet is protosolar. These authors also limit the possible values of the WMF to 77\% at maximum, according to measurements on cometary compositions. In our case, we lower this value to 50\%, based on our knowledge of the interiors of large icy satellites such as Titan \citep{Tobie06}. The areas on the ternary diagram corresponding to these restrictions are thus shaded (see Figure~\ref{fig:3}). Interestingly, Mercury lies inside this exclusion region but its current state could result from a post-formation alteration such as mantle evaporation due to strong melting \citep{Cameron85} or a giant impact during the early phases of its evolution \citep{Benz88}. Moreover, one must not forget that the planets considered in this work are made from a limited number of materials. Planets harbouring a thick gaseous atmosphere can easily be larger than the maximum value allowed with water, however we consider in this study only dense solid planets (with possible addition of liquid water) without thick gaseous atmospheres.

\begin{figure}[h]
\begin{center}
\includegraphics[width=0.9\columnwidth]{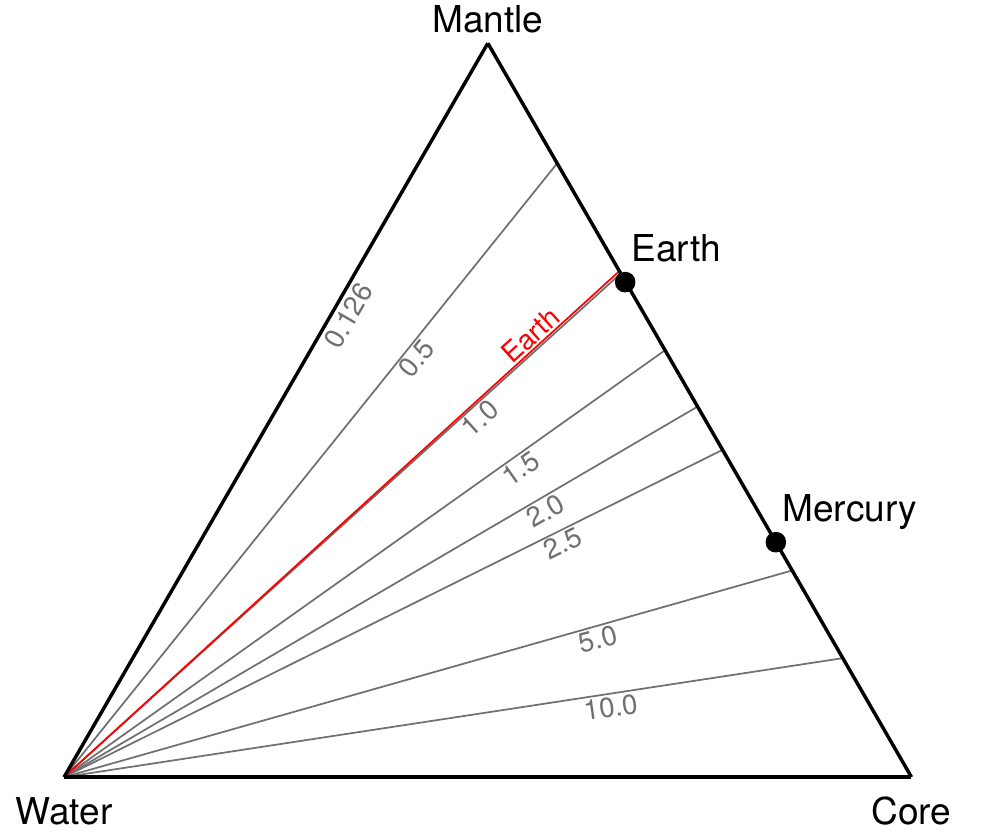}
\caption{Isolines of constant planetary Fe/Si ratio in the ternary diagram. The Earth's value is shown in red.}
\label{fig:4}
\end{center}
\end{figure}

These restrictions on the ternary diagram are based on our current knowledge of the solar system bodies, and they may not be suitable for application to all exoplanetary systems, as they could have had different planet formation conditions. Morever, these restrictions do not break the degeneracy existing on a planet's composition. This would however be possible if the Fe/Si ratio of the planet is known, since it is strongly related to the CMF value. With our assumptions, the Fe/Si ratio of a planet is independent from its mass, and is essentially governed by the CMF and WMF (see Appendix). Thus, as for the radius, we can draw isolines of constant Fe/Si ratios in the ternary diagram (see Figure~\ref{fig:4}). The intersection of the isoline of a planet's Fe/Si with its isoradius curve would then be the only composition allowed for this planet. The Fe/Si and Mg/Si bulk ratios of an exoplanet cannot be directly measured but their host star's values can provide a good approximation \citep{Thiabaud15}. Here, we assume that the Fe/Si ratios of the exoplanet and its host star are similar to break the degeneracy on the planet's composition.

\section{Results}
\label{sec:3}

\subsection{Mass-radius relationships}
\label{ssec:3.1}

\begin{figure}
\begin{center}
\includegraphics[width=\columnwidth]{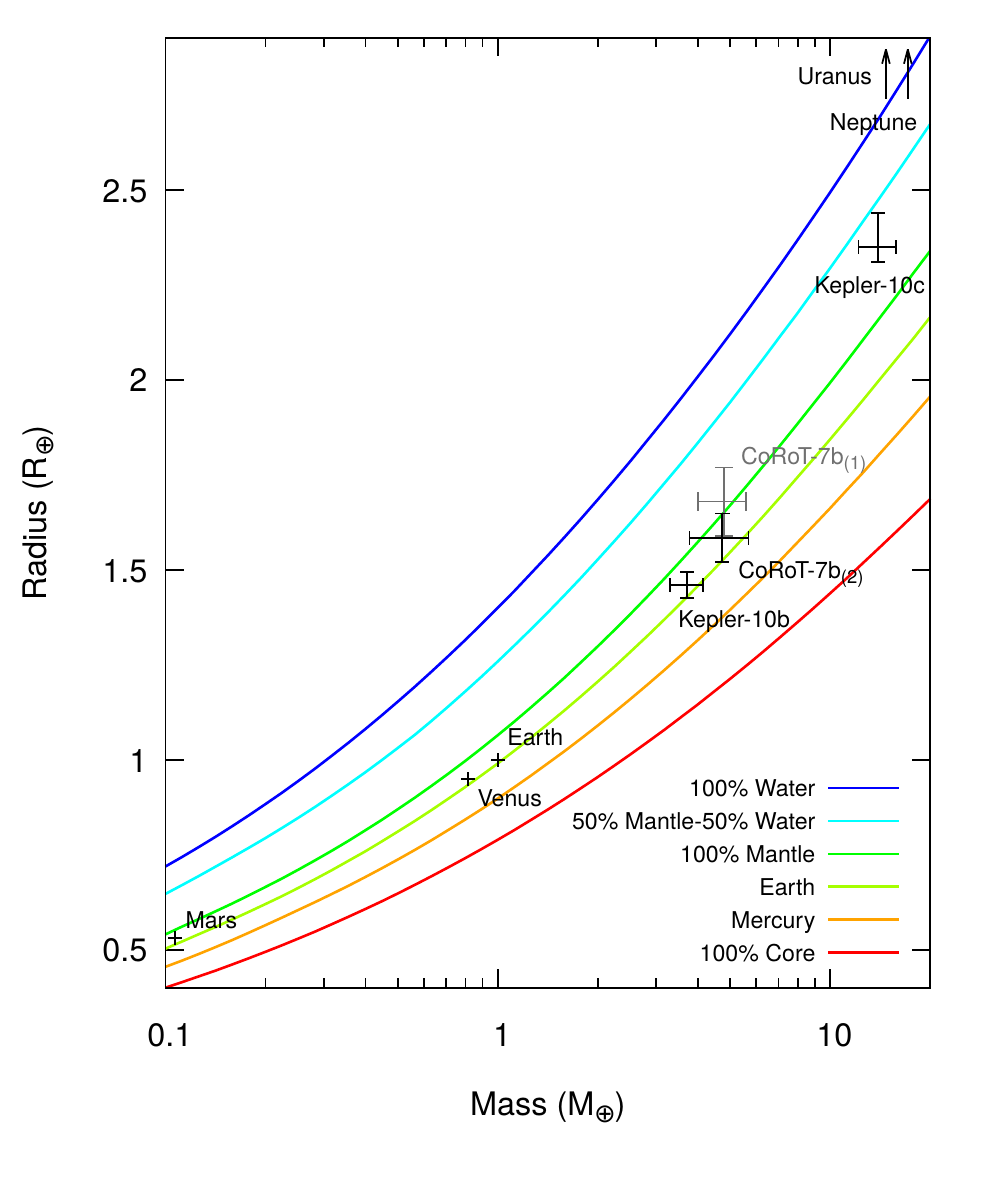}
\caption{From top to bottom, mass-radius curves for different planet compositions: 100\% water, 50\% mantle--50\% water, 100\% mantle, Earth-like, Mercury-like, and 100\% core. Solar system planets that have a mass in the explored range are shown. Three exoplanets, with the measurement errors on their physical parameters, are also shown: CoRoT-7b with original (1) and updated (2) values (see Section~\ref{ssec:3.2}), Kepler-10b, and Kepler-10c \citep{Dumusque14,Weiss16}. The data used for this figure are available online (see text).}
\label{fig:5}
\end{center}
\end{figure}

Our model allows the computation of $R_{\rm P}$ as a function of $M_{\rm P}$ for different compositions. Comparing these values to the measured physical parameters then provides a first estimate of the exoplanet's composition. Here we chose six typical compositions, namely 100\% water, 50\% mantle--50\% water, 100\% mantle, Earth-like, Mercury-like, and 100\% core (see Figure~\ref{fig:5}). As explained in Section~\ref{ssec:2.6}, planets with extreme compositions like 100\% water or 100\% core are unlikely to form, however the mass-radius curves corresponding to these compositions provide good theoretical markers for the validation of fundamental parameters of detected exoplanets. If an exoplanet is located under the 100\% core curve, it challenges its mass and radius estimates, as no planetary body can form with a density higher than pure iron. On the other hand, a planet located beyond the 100\% water curve cannot be composed of the materials used here, and most likely harbours a significant gaseous atmosphere. The parameters $f_{\rm alloy}$, $\left( \frac{\rm Mg}{\rm Si} \right) _{\rm P}$, and $\rm Mg\#$ are taken equal to the Earth's value (see Table~\ref{tab:1}). We explore planetary masses up to 20~$\mearth$, which corresponds to the upper limit of validity of the Vinet EOS (pressures of the order of 1--10~TPa; \citealp{HamaSuito96}). Placing some well-known exoplanets on Figure~\ref{fig:5} provides indications of their possible compositions. Kepler-10c, a $\sim$14~$\mearth$ planet \citep{Weiss16}, lies above the line corresponding to a 100\% mantle composition, i.e. the least dense fully rocky composition, and probably harbours a significant fraction of water. On the other hand, planets like Kepler-10b and CoRoT-7b, with the latest estimates of their fundamental parameters, are compatible with fully rocky compositions, and even with an Earth-like composition. A more detailed study of these exoplanets is provided in Section~\ref{ssec:3.2}.

Previous studies found that, for an Earth-like composition, the planet radius grows proportional to $M_{\rm P}^{0.274}$ \citep{Sotin07} or $M_{\rm P}^\text{0.267--0.272}$ \citep{Valencia06} for $M_{\rm P} \in$ [1--10]$\mearth$. When considering ocean planets (with a 50\% WMF) the power exponent remains similar, but the radius increases faster ($R_{\rm P} = 1.262$~$M_{\rm P}^{0.275}$; \citealp{Sotin07}) because of the lower density of the water phases. For a Mercury-like composition, the power exponent becomes $\sim$0.3 \citep{Valencia06}. Here, when fitting our mass-radius curves, we obtain $R_{\rm P}~=~1.00$~$M_{\rm P}^{0.270}$ for an Earth-like composition and $M_{\rm P}~\in$~[1--10]~$\mearth$. The power exponent is lowered for ocean planets (50\% mantle--50\% water) ($R_{\rm P}~=~1.27$~$M_{\rm P}^{0.261}$) and for Mercury-like planets ($R_{\rm P}~=~0.90$~$M_{\rm P}^{0.268}$). The value close to 0.3 found by \cite{Valencia06} is only retrieved within the 0.1--1~$\mearth$ range. Overall, we obtain lower values compared to \cite{Valencia06} and \cite{Sotin07}. This discrepancy originates from the extrapolation differences between the Vinet EOS and the BM3 EOS, since the latter results in an overestimation of the planet radius. Because the power laws detailed here do not provide perfect fits of the mass-radius curves represented in Figure~\ref{fig:5}, we recommend the use of the machine-readable table available online.

\subsection{Compositions of Super-Earths}
\label{ssec:3.2}

In the following, when investigating the composition of an exoplanet without inclusion of water, we draw a colormap of the computed planet radii as a function of composition and of the planet's measured mass range. In the case where we consider the presence of water, we draw instead three ternary diagrams corresponding to the minimum ($M_{\rm P}-\delta M_{\rm P}$), central ($M_{\rm P}$), and maximum ($M_{\rm P}+\delta M_{\rm P}$) values of its mass range, as a complete investigation of this range cannot be represented in two dimensions. To explore the impact of the uncertainties $\delta R_{\rm P}$ on the planet radius $R_{\rm P}$, we draw the three isoradius curves $R_{\rm P}-\delta R_{\rm P}$, $R_{\rm P}$, and $R_{\rm P}+\delta R_{\rm P}$ on each diagram. On one diagram, the domain included within the $R_{\rm P}-\delta R_{\rm P}$ and $R_{\rm P}+\delta R_{\rm P}$ curves corresponds to the set of compositions allowed for this planet, and for the considered mass in the case of a ternary diagram. We further reduce this set of compositions by considering the Fe/Si ratio of the planet's host star. From this data, we provide the ranges of plausible values for the CMF and WMF allowed in the planet, considering the uncertainties on $M_{\rm P}$, $R_{\rm P}$, and on the stellar Fe/Si ratio, assuming that the planet does not harbour a thick gaseous atmosphere.

\subsubsection{CoRoT-7b}

CoRoT-7b is the first detected Super-Earth with known mass and radius. The discovery of this planet was reported by \cite{Leger09} who found a radius of 1.68~$\pm$~0.09~$\rearth$ and an orbital period of 0.85359~$\pm$~5~10$^{-5}$~day. Its mass was obtained shortly after by \cite{Queloz09}, who derived a value of 4.8~$\pm$~0.8~$\mearth$ using radial velocity measurements, and bestowed the Super-Earth status to CoRoT-7b from its derived mean density. Meanwhile, \cite{Valencia10} performed a detailed study of CoRoT-7b's interior and composition, as well as of the mass loss resulting from the proximity of the planet to its host star (0.0172~$\pm$~0.00029~AU; \citealp{Queloz09}). They concluded that the planet could not retain a gaseous atmosphere made of H/He because of stellar irradiation, implying the exclusion of such a layer. However, they found that the values of $M_{\rm P}$ and $R_{\rm P}$ could still be consistent with a solid planet, provided that its iron content is significantly depleted compared to the Earth. An Earth-like composition was only reached with a 1$\sigma$ increase in mass ($M_{\rm P} = 5.6$~$\mearth$) and a 1$\sigma$ decrease in radius ($R_{\rm P} = 1.59$~$\rearth$) compared to the central values.

The physical parameters of CoRoT-7b have been refined by \cite{Barros14} and \cite{Haywood14}, updating the values of radius and mass to 1.585~$\pm$~0.064~$\rearth$ and 4.73~$\pm$~0.95~$\mearth$, respectively. Using these new values and assuming that the planet is only made of solid materials (with the possible addition of liquid water), we have re-evaluated here the internal structure of CoRoT-7b. For this, we have explored the domain of planetary compositions corresponding to one of CoRoT-7b's possible formation scenarios, i.e. \textit{in situ} formation (or formation close to the star). Two of the compositional parameters that are not represented in our diagrams ($f_{\rm alloy}$ and $\rm Mg\#$) are by default taken equal to the Earth's values. The surface temperature of the planet is taken equal to the estimated equilibrium temperature of CoRoT-7b (1756~$\pm$~27~K; \citealp{Barros14}), whereas the surface pressure is set to 1~bar to mimic the presence of a light atmosphere.

The Fe, Mg and Si elemental abundances in CoRoT-7 have been derived from high resolution spectroscopy by \cite{Bruntt10}. These authors also determined the abundances of Al, Ca, and Ni, among many other refractory elements, in CoRoT-7. Here, following the approach of \cite{Sotin07}, these three elements are added to Fe, Mg, and Si and used to correct the Fe/Si and Mg/Si bulk ratios employed in the model (see Table~\ref{tab:1}). In the case of the Earth, Fe, Mg, Si, O, and S only account for 95\% of the planet's mass, whereas the addition of Al, Ca, and Ni enlarges this fraction to more than 99\% \citep{MorganAnders80,Allegre95}. From these considerations, we get corrected Fe/Si and Mg/Si ratios equal to 0.826~$\pm$~0.419 and 1.036~$\pm$~0.614 for the bulk composition of CoRoT-7b, respectively. Contrary to the Fe/Si ratio whose full range is investigated through the model, only the central Mg/Si value is used as an input parameter because this latter does not significantly impact the computed radius \citep{Sotin07}.

\begin{figure}
\begin{center}
\includegraphics[width=\columnwidth]{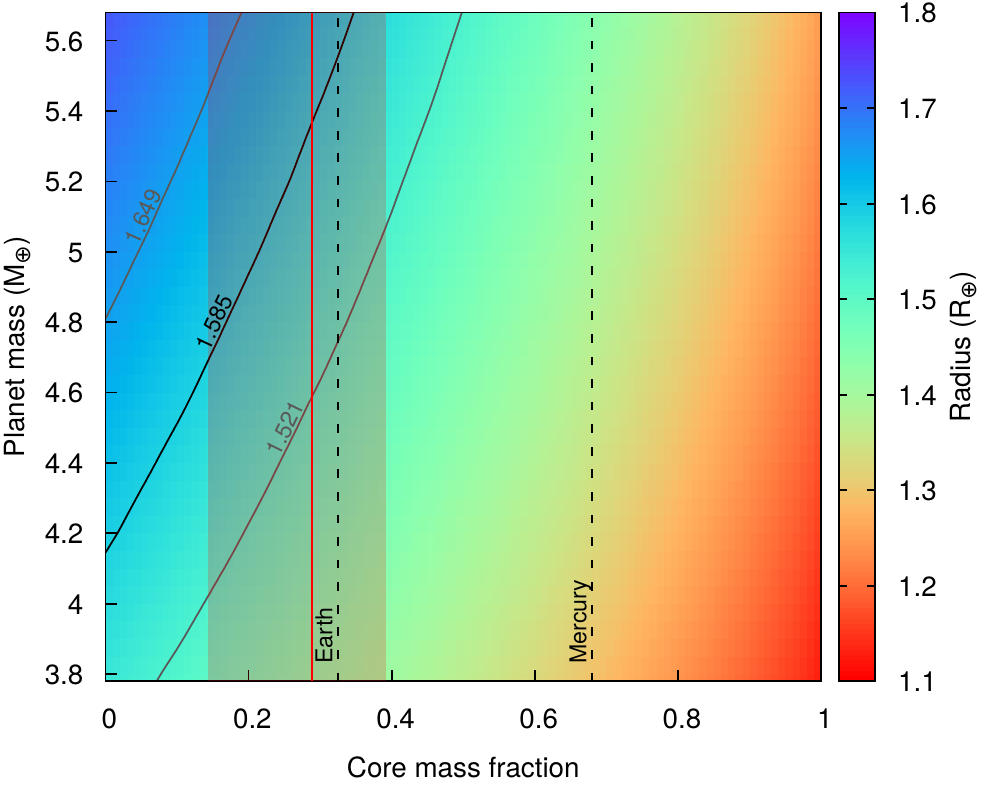}
\caption{Computed planet radii for CoRoT-7b as a function of composition (X-axis, corresponding to the CMF as we assume WMF=0) and mass (Y-axis, from the 1$\sigma$-range inferred by \cite{Haywood14}). Also shown are the isoradius curves denoting the planet radius measured by \cite{Barros14} with the 1$\sigma$ extreme values. The Fe/Si ratio assumed for CoRoT-7b, with its associated uncertainties, delimit a line and an area represented in red.}
\label{fig:6}
\end{center}
\end{figure}

As for the terrestrial planets in the solar system, CoRoT-7b may have formed \textit{in situ}, or at least inside the snow line of the protoplanetary disk that surrounded the host star. In this region, the temperature in the protoplanetary disk was too high for water to condense, leading to formation of fully rocky planets. Figure~\ref{fig:6} shows the result of the exploration of CoRoT-7b's compositional parameter space, if we assume completely dry compositions only (i.e., WMF=0). It illustrates the fact that an Earth-like composition is easier to achieve than with the previous planet's parameters. This particular composition can be obtained for all planetary masses between 4.73 and 5.68~$\mearth$, considering the 1$\sigma$ uncertainty on the planet's radius. The conclusion of \cite{Valencia10}, i.e. the fact that CoRoT-7b needs to be depleted in iron to explain the measured parameters, is no longer required with the updated parameters, since now the planet can present a CMF as large as that of the Earth. Continuing with the assumption of a dry composition, we show that CoRoT-7b's CMF may vary between 0 and 50\%, depending on the uncertainties on the fundamental parameters. However, when our estimate of the Fe/Si ratio is considered in the planet, this range becomes limited to 13--37\%.

\subsubsection{Kepler-10b}

\begin{figure}
\begin{center}
\includegraphics[width=\columnwidth]{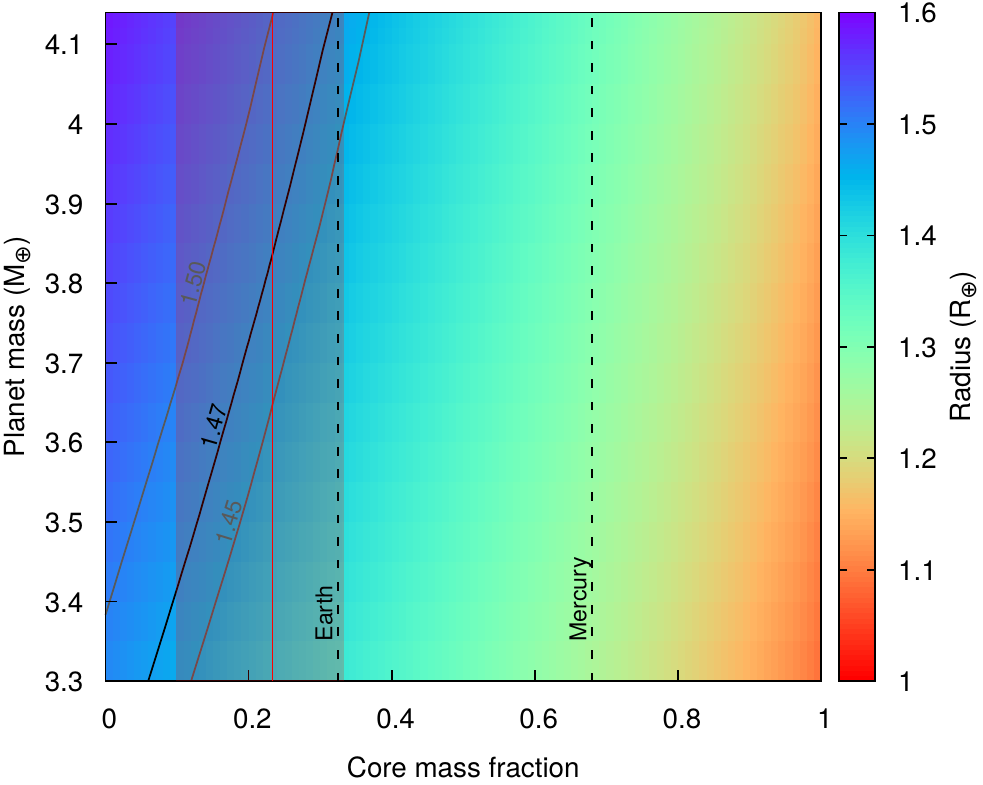}
\caption{Computed planet radii for Kepler-10b as a function of composition (X-axis, corresponding to the CMF as we assume WMF=0) and mass (Y-axis, from the 1$\sigma$-range inferred by \cite{Weiss16}). Also shown are the isoradius curves denoting the planet radius measured by \cite{Dumusque14} with the 1$\sigma$ extreme values. The Fe/Si ratio assumed for Kepler-10b, with its associated uncertainties, delimit a line and an area represented in red.}
\label{fig:7}
\end{center}
\end{figure}

Kepler-10b is the first rocky planet that has been detected by the \textit{Kepler} mission \citep{Batalha11}. While it is in many aspects comparable to CoRoT-7b, its radius ($1.47 ^{+0.03}_{-0.02}$~$\rearth$; \cite{Dumusque14}) and mass (3.72~$\pm$~0.42~$\mearth$; \cite{Weiss16}) were measured with a better precision. \cite{Weiss16} performed a study of Kepler-10b's interior, showing that a fully rocky composition is compatible with the measurements of mass and radius. They also derived a CMF of the planet within the 0.17~$\pm$~0.12 range in the solid case. Here, we use the host star's chemical composition derived by \cite{Santos15}, and compute the corresponding Fe/Si and Mg/Si ratios. Unlike the case of CoRoT-7, the Al, Ca, and Ni abundances have not been measured in Kepler-10. We then find Fe/Si = 0.708~$\pm$~0.375 and Mg/Si = 1.230~$\pm$~0.595. Figure~\ref{fig:7} shows the improved precision on Kepler-10b's radius, as the domain of possible compositions delimited by the 1$\sigma$ isoradius lines is more restrained for this planet than for CoRoT-7b.

As for CoRoT-7b, we investigate the possibility that Kepler-10b formed inside the snow line (see Figure~\ref{fig:7}). In the case of a fully rocky planet with $M_{\rm P} = 3.72$~$\mearth$, the CMF of Kepler-10b varies between 12 and 26\%, namely roughly within the upper half of the range of \cite{Weiss16}. With a 1$\sigma$ decrease on the planet's mass, this range becomes 0--13\%. On the other hand, with a 1$\sigma$ increase of the mass, the CMF becomes 24--38\%, a range of values becoming closer to those estimated for the Earth or Venus. In particular, a fully rocky Kepler-10b with $M_{\rm P} = 4.14$~$\mearth$ and $R_{\rm P} = 1.47$~$\rearth$ presents the same CMF as for the Earth (32\%). When considering the limitations imposed by the Fe/Si ratio, the CMF associated to $\pm$1$\sigma$ in mass is now reduced to 10--13\% and 24--33\%, respectively. The overall upper limit on the CMF is thus much closer to the Earth's value. The range evaluated for the central mass is not affected by the consideration of the planet's Fe/Si ratio. 

We find the possible CMF range of Kepler-10b to be in the range 10--33\%, which is in good agreement with the results from \cite{Weiss16}. However, their results ruled out an Earth-like composition for Kepler-10b, whereas we show here that this composition is possible. Kepler-10b appears to be one of the best cases for the study of exoplanetary composition, thanks to both the high precision on its mass and radius and the measurement of its host star's elemental abundances. Together, these two features reduce the set of compositions allowed for this planet.

\subsubsection{Proxima Centauri b}

Proxima Centauri b, a low-mass planet orbiting the Sun's closest neighbor, was recently discovered by \cite{AngladaEscude16}. The planet's radius remains unknown because no transit has been detected so far \citep{Kipping16}, and the only known physical parameter is the planet's minimum mass $M_{\rm P}$sin~$i$ found to be $1.27 ^{+0.19}_{-0.17}$~$\mearth$ \citep{AngladaEscude16}. \cite{Brugger16} have performed a study of the possible interiors and compositions of Proxima b as a rocky body, with possible addition of water, assuming that sin~$i = 1$. In their study, the computed radius of Proxima b spans the 0.94--1.40~$\rearth$ range, the minimum value being obtained in the case of a 1.10~$\mearth$ dry planet with a 65\% CMF, and a maximum value reached when considering $M_{\rm P} = 1.46$~$\mearth$ for an ocean planet with 50\% water in mass. Here, we extend the study of \cite{Brugger16} by exploring the impact of a large range of values of sin~$i$ on the mass of Proxima b, still assuming that the planet does not possess a thick gaseous atmosphere. From our computations, there is a 77\% probability that the planet's mass is within the range [1.27--2]~$\mearth$, which we find to correspond to a maximum orbital inclination of 39.4\degree. If the mass range is extended to [1.27--5]~$\mearth$ (i.e. $i = 14.7\degree$), the computed probability reaches 96.7\%. Interestingly, Proxima b is located on a temperate orbit around its host star, with an equilibrium temperature estimated at 234~K \citep{AngladaEscude16}. Therefore, liquid water can easily be stable on its surface with the presence of an Earth-like, light atmosphere, and supercritical water is not needed to model the current state of the planet, in contrast to CoRoT-7b and Kepler-10b.

\begin{figure}
\begin{center}
\includegraphics[width=0.9\columnwidth]{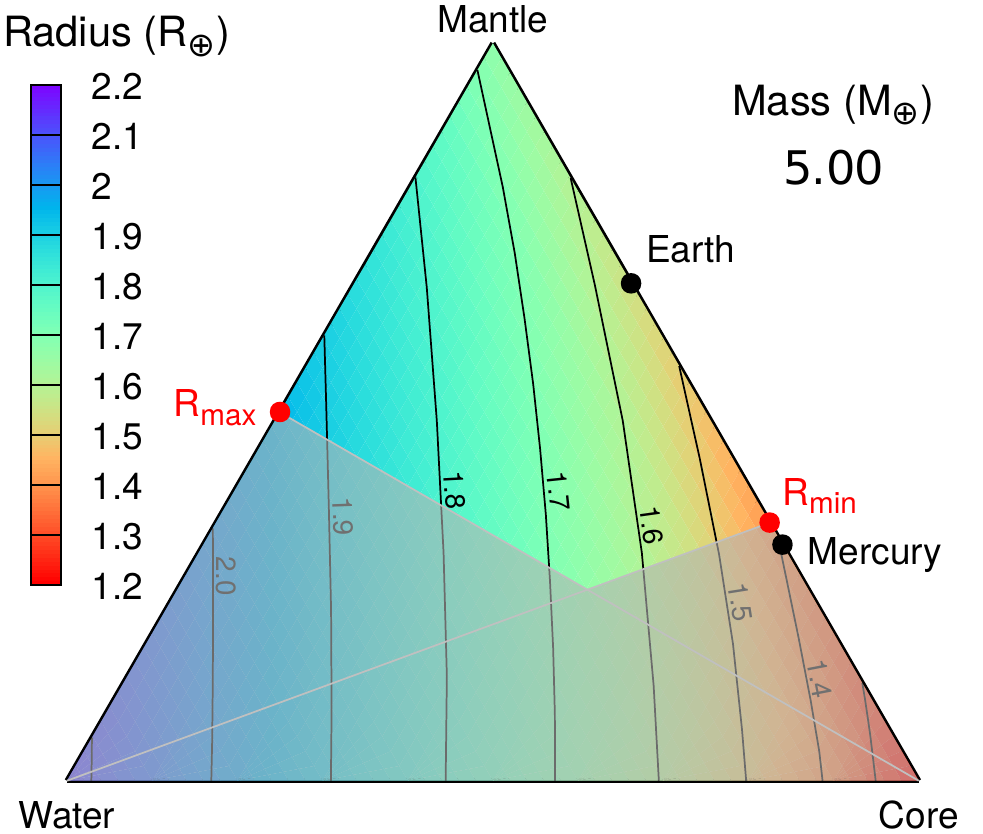}
\caption{Ternary diagram of all compositions explored for a 5~$\mearth$ Proxima b, with the minimum and maximum computed planet radii corresponding to compositions A and B (red dots, see text), and the isoradius curves from 1.3 to 2.1~$\rearth$ (black curves).}
\label{fig:8}
\end{center}
\end{figure}

Let ``A'' and ``B'' be the compositions that produce the minimum and maximum radii for a given value of $M_{\rm P}$, namely (CMF,WMF) = (0.65,0) and (0,0.5), respectively, according to the characteristics of solar system bodies (see Section~\ref{ssec:2.6}). With $M_{\rm P} = 2$~$\mearth$, we obtain $R_{\rm min} = 1.10$~$\rearth$ for composition A, and $R_{\rm max} = 1.53$~$\rearth$ for composition B. These values do not differ significantly from the results found by \cite{Brugger16}. The planet's radius computed for $M_{\rm P} = 5$~$\mearth$ spans the range 1.41--1.94~$\rearth$ (see Figure~\ref{fig:8}). These results are in good agreement with the density-radius relationship obtained by \cite{Weiss16} from statistical analysis of Kepler planets with known masses and radii, in the regime of planets smaller than 4~$\rearth$. These authors derive two regimes in which the mean density increases with the radius up to 1.5~$\rearth$ and then decreases for radii in the 1.5--4~$\rearth$ range. The maximum mean density is thus found to be 7.6~g/cm$^3$ with a corresponding radius of 1.5~$\rearth$. In our simulations, the case of a 5~$\mearth$ Proxima b with composition A yields a mean density of 9.9~g/cm$^3$, significantly higher than this statistical maximum value. In the case of composition B, we obtain $M_{\rm P} = 5$~$\mearth$ and $R_{\rm P} = 1.94$~$\rearth$, giving a mean density of 3.8~g/cm$^3$, which agrees well with the relationship found by \cite{Weiss16}. The capability of Proxima b to retain a possible liquid water ocean facing the strong stellar irradiation is uncertain \citep{Ribas16}. However \cite{Airapetian17} recently concluded that the planet cannot be habitable, since an Earth-like atmosphere would escape in $\sim$10~Myr because of the star's strong XUV flux. In the case of Proxima b being a Super-Earth with a mass up to 5~$\mearth$, the higher escape velocity at its surface would not be sufficient to retain the atmosphere, as this effect would be counterbalanced by a weaker magnetic field, compared to an Earth-sized planet \citep{Airapetian17}.

\section{Discussion and Conclusions}
\label{sec:4}

We have developed an internal structure model coupled to a numerical scheme that explores the compositional parameter space of a planet. In our approach, we have assumed that the Fe/Si ratio measured in host stars is similar to that of their orbiting planets, allowing us to significantly decrease the degeneracy on their compositions. Applying this model to CoRoT-7b and Kepler-10b with the latest estimates of their mass and radius, we show that both planets present CMF values varying within the 10--37\% and 10--33\% ranges, respectively. The fundamental parameters of these two planets are consistent with fully rocky compositions, and especially with an Earth-like composition. These results are compatible with a formation inside the snow line for both planets, are are in good agreement with those of \cite{Dorn17} who studied the interiors of those planets using a bayesian method, and also found that they are compatible with an Earth-like composition. They showed that the CMF of Kepler-10b's and CoRoT-7b spans a range approximately between 0 and that of the Earth. Note that we have modeled these planets assuming surface temperatures equal to their equilibrium temperatures with surface materials in solid phases. Such high surface temperatures should however lead to the formation of melted silicates oceans \citep{SchaeferFegley09}, a phase whose density differs from those of the solid phases by a few percent \citep{Leger11,Lebrun13}. Reporting this difference in density in our model produces a difference on the computed radius by less than 1\%. However, the impact of melted silicates on a planet's radius could be more important, as other material properties, namely the bulk modulus, may be strongly altered. In addition, the heat transfer in this layer would be modified compared to solid materials. This would require, in particular, a finer description of the planet's thermal profile, which is not investigated here.

Alternatively, these planets may have formed further away from their host star. In that case, water would have condensed along with silicate and metal grains, forming planets that contain a significant amount of water in liquid or solid phases. To reach their current orbital positions, water-rich CoRoT-7b and Kepler-10b would have migrated inward from their formation region. Moreover, given their current equilibrium temperature, water on their surface would have turned into vapor and supercritical phases (for $T_{\rm surf} > 647$~K; \citealp{WagnerPruss02}) instead of liquid.

Finally, we have considered the possibility that the recently discovered Proxima Centauri b presents a mass reaching 5~$\mearth$ (which covers 96.7\% of the orbital inclinations of the system), and shown that its radius may be as large as 1.94~$\rearth$ when considering an important amount of volatiles (50\% water in mass). This value is still below the mean mass-radius relationship found by \cite{Weiss16} for low-mass terrestrial exoplanets.

A third formation scenario has been proposed by \cite{Lee14}. This model assumes a formation below the snow line with accretion of nebular gas, even for low-mass planets, which could thus harbour an H$_2$/He atmosphere of a few percent in mass. In the case of CoRoT-7b, \cite{Valencia10} showed that stellar irradiation is too strong for the planet to retain such an atmosphere. This also applies for Kepler-10b, which undergoes similar conditions, with an orbital distance and a mass even smaller than those of CoRoT-7b \citep{Dumusque14,Weiss16}. This third formation scenario may apply to Proxima b, which could still harbour a light primary atmosphere, especially given its unknown actual mass. Modeling the interior of a gaseous Proxima b is thus needed to complete the investigation of this planet. Note however that giant gaseous planets appear to be relatively rare at short orbits around M dwarf stars \citep{Bonfils13,Tuomi14,DressingCharbonneau15}.

\subsection{Diversity of planetary materials}
\label{ssec:4.1}

Since our model is predominantly based on the Earth's interior, the existence of the materials considered here is only verified within ranges of pressure ($10^5$--$10^{11}$~Pa) and temperature (300--6000~K) that reign inside our planet. Outside these ranges, the used materials transform into different phases that should be present in the case of the investigated extrasolar planets. At the bottom of the Earth's mantle, where the pressure values are $\sim$120~GPa, the existence of a post-perovskite phase ($\sim$1.5\% denser than bridgmanite) has been observed \citep{OganovOno04,Murakami04} and incorporated by \cite{ZengSasselov13}. They also considered another transformation of this new phase around 1~TPa. The impact on their results is however small compared to simpler models.

The budget of light elements in the Earth's core is still a matter of discussion. In addition to sulfur, the outer core may contain oxygen and silicate up to 5\% and 3.6\%, respectively \citep{Badro15}. If a substantial amount of light element (in addition to sulfur) also enters the composition of super-Earths' cores, these cores may be less dense than assumed in our models. As a consequence, the fraction of water required to explain the observed mass of super-Earth may be significantly reduced. The presence of silicon in particular would add a level of degeneracy to the composition of studied planets, as the value of the bulk Fe/Si ratio would not be enough to set the planet's core mass fraction. Therefore, the impact of the presence of volatiles on the computation of a planet's radius has to be studied thoroughly, to determine if this parameter has a significant effect on the radius of exoplanets.

At the top of the mantle, on the other hand, olivine and pyroxenes may not be stable if the thickness of the outer ice layer is such that the pressure at its bottom exceeds 25~GPa \citep{Irifune87}. This would leave a mantle entirely composed of bridgmanite and ferro-periclase. Alternatively, if pressure is slightly lower than 25~GPa, an upper mantle may exist but would be very thin. Convection within this layer would be weak or may not operate at all, which would strongly alter the cooling and thermal state of such ocean planets. Here, for CoRoT-7b and Kepler-10b, we compute that the upper mantle is no longer present for WMF approaching $\sim$10\%. A 5\% WMF yields a $\sim$160~km thick upper mantle in both cases, whereas this value goes down to $\sim$15~km for 8\% water in mass. In comparison, the Earth's upper mantle is about 700~km thick. In addition, as was proposed for icy moons of the solar system \citep{Deschamps10}, ice layers may be composed of a mix of water and volatile compounds by a few percent in mass. While the presence of volatiles may only slightly affect the density of the water ice VII layer (and thus the mass-radius relationships), it may change physical properties more substantially, in particular thermal conductivity \citep{HsiehDeschamps15}, and may add to the alteration of heat transfer from the presence of a thin upper mantle.

As for silicate rocks, water ice presents high-pressure phases that have been measured through laboratory experiments, such as the ice VII--ice X transition around 70~GPa, ice X having a behavior still close to that of ice VII \citep{Sotin07}. Above these pressures, new phases have been predicted by theory around 1~TPa \citep{MilitzerWilson10}. These pressures cannot be reached in the case of ocean planets up to 10~$\mearth$ (the value reached at the bottom of the ice VII layer for these planets is $\sim$400~GPa). However they may appear in the case of icy giant planets such as Uranus or Neptune.

Exoplanets may also be formed from materials different from those encountered in our own solar system. In particular, the silicate rocks we consider are based on oxygen, as our own protosolar nebula had a C/O ratio of $\sim$0.54 \citep{Asplund09}. However, a small fraction of stars present higher C/O ratios \citep{BrewerFischer16}, suggesting the possibility of forming carbon-rich planets around such stars. For example, \cite{Madhusudhan12} studied the interior of a possible carbon-rich 55 Cancri e, with the incorporation of pure carbon and silicon carbide, given the high C/O ratio of its host star 55 Cancri. Even if this latter ratio has been re-estimated at a lower value (0.78~$\pm$~0.08; \citealp{Teske13}), the existence of carbon-rich planets opens another branch of exoplanetary science.

\subsection{Prospects}
\label{ssec:4.2}

As observed in Section~\ref{ssec:2.5}, modeling the interior of exoplanets with three major main layers (here core, mantle, and water) generates a degeneracy on their relative fractions, as different sets of these fractions produce the same planet radius. Adding layers made of another material, such as an H/He atmosphere, would thus add a degree of degeneracy, as the isoradius curves in the compositional parameter space would become isoradius surfaces (the four-variables equivalent of the ternary diagram being a tetrahedron). Therefore, there is an important need to break the degeneracy existing with solid planets. We have shown here that this degeneracy can be significantly reduced under the assumption that the stellar Fe/Si and Mg/Si ratios are similar to those of the planet \citep{Thiabaud15}. However, the planet-hosting stars whose ratios are known from high-resolution spectroscopy represents a small fraction of the $\sim$3500 currently confirmed planetary systems\footnote{exoplanetarchive.ipac.caltech.edu}. There is thus an important need to measure the elemental abundances of these stars along with the physical and orbital parameters of the systems. By coupling these measurements to the new generation of space-born observatories such as PLATO or CHEOPS, which will supply physical and orbital parameters of exoplanets around bright stars with an unrivaled precision, we should be able to significantly reduce the set of compositions allowed for a given exoplanet, and thus provide with precise ranges of its core and water mass fractions.

\acknowledgements
The project leading to this publication has received funding from Excellence Initiative of Aix-Marseille University - A*MIDEX, a French ``Investissements d'Avenir'' program. O.M. and M. D. also acknowledge support from CNES. We thank an anonymous referee for her/his very constructive comments that helped us strengthen our manuscript.

\appendix

\section{Supplementary material}

\subsection{Detailed composition of a planet}

To fix the distribution of chemical species in the different layers of the planet (e.g. Fe and FeS in the core, and the different silicate rocks in the mantles; see Section~\ref{ssec:2.1}), we introduce the mole fractions $x_i$ and $y_i$ (with $i \in \llbracket 1,5 \rrbracket$ the number of the layer), as detailed in Table~\ref{tab:3}. The use of two variables allows to manage two levels of distribution.

\begin{deluxetable}{lcccccccccccc}
	\tablecaption{Distribution of the different materials (level 1) and chemical species (level 2) in the five layers}
	\tablehead{
		\colhead{Layer}	& \multicolumn{2}{c}{1. Core}	& \multicolumn{4}{c}{2. Lower mantle}	& \multicolumn{4}{c}{3. Upper mantle}	& \colhead{4. Ice VII}	& \colhead{5. Liquid water}
	}
	\startdata
		\multirow{2}{*}{Level 1}		& \multicolumn{2}{c}{$100\%$}		& \multicolumn{2}{c}{$x_2$}		& \multicolumn{2}{c}{$1-x_2$}		& \multicolumn{2}{c}{$x_3$}		& \multicolumn{2}{c}{$1-x_3$}		& $100\%$		& $100\%$		\\
				& \multicolumn{2}{c}{Metal}			& \multicolumn{2}{c}{Bridgmanite}		& \multicolumn{2}{c}{Periclase}		& \multicolumn{2}{c}{Olivine}		& \multicolumn{2}{c}{Enstatite}		& Ice VII		& Liq. H$_{2}$O		\\
		\hline
		\multirow{2}{*}{Level 2}		& $y_1$		& $1-y_1$		& $y_2$		& $1-y_2$		& $y_2$		& $1-y_2$		& $y_3$		& $1-y_3$		& $y_3$		& $1-y_3$		& \multirow{2}{*}{$\varnothing$}		& \multirow{2}{*}{$\varnothing$}		\\
				& FeS		& Fe			& FeSiO$_{3}$		& MgSiO$_{3}$		& FeO		& MgO		& Fe$_{2}$SiO$_{4}$		& Mg$_{2}$SiO$_{4}$		& Fe$_{2}$Si$_{2}$O$_{6}$		& Mg$_{2}$Si$_{2}$O$_{6}$		& 		& 		\\
	\enddata
	\label{tab:3}
\end{deluxetable}

The mantle is assumed to be chemically homogeneous, meaning that the Fe/Si and Mg/Si mole ratios are identical in the lower and upper parts of the mantle (respectively layers 2 and 3). If $\left( \frac{\rm X}{\rm Y} \right) _i$ is the value of the X/Y mole ratio in the layer $i$, then:

\begin{equation}
	\left( \frac{\rm Fe}{\rm Mg} \right) _2 = \left( \frac{\rm Fe}{\rm Mg} \right) _3 \qquad \Longleftrightarrow \qquad y_2 = y_3
	\label{eq:1}
\end{equation}
\begin{equation}
	\left( \frac{\rm Mg}{\rm Si} \right) _2 = \left( \frac{\rm Mg}{\rm Si} \right) _3 \qquad \Longrightarrow \qquad x_2 = 1 - \frac{x_3}{2}
	\label{eq:2}
\end{equation}

To link the mole fractions listed in Table~\ref{tab:3} to the composition of a planet, we define the following parameters:

\begin{itemize}
	\item $f_{\rm alloy}$ the fraction of iron alloy in the core;
	\item $\left( \frac{\rm Mg}{\rm Si} \right) _{\rm P}$ the planet's overall Mg/Si mole ratio;
	\item $\rm Mg\# \equiv \left( \frac{\text{Mg}}{\text{Mg}+\text{Fe}} \right) _{Mantle}$ the Mg number \citep{Sotin07}.
\end{itemize}

The Mg number $\rm Mg\#$, which reflects the amount of iron in the mantle, expresses the degree of differentiation of a planet. For instance, the Earth ($\rm Mg\# = 0.9$; \citealp{Sotin07}) is more differentiated than Mars ($\rm Mg\# = 0.7$; \citealp{Sotin07}) due to their difference in mass, Mars being ten times less massive.

From Equations~\ref{eq:1}~and~\ref{eq:2}, and following the definitions of the three aforementioned parameters, we obtain:

\begin{equation*}
	\begin{cases}
      x_1 = 1																	\\
      x_2 = \rm Mg\# / \left( \frac{\rm Mg}{\rm Si} \right) _{\rm P}			\\
      x_3 = 2(1-x_2)															\\
      x_4 = 1																	\\
      x_5 = 1
    \end{cases}
    \qquad
    \qquad
    \text{and}
    \qquad
    \qquad
    \begin{cases}
      y_1 = f_{\rm alloy}														\\
      y_2 = 1 - \rm Mg\#														\\
      y_3 = y_2																	\\
      y_4 = 0																	\\
      y_5 = 0
    \end{cases}
\end{equation*}

Thus, the internal structure of a planet can be entirely described by six compositional parameters: the planet's mass $M_{\rm P}$, the core mass fraction (CMF) $x_{\rm core}$ and water mass fraction (WMF) $x_{\rm water}$, the fraction of iron alloy in the core $f_{\rm alloy}$, the Mg/Si mole ratio of the planet $\left( \frac{\rm Mg}{\rm Si} \right) _{\rm P}$, and the Mg number $\rm Mg\#$.

\subsection{Equations}

Once the composition of a planet is fixed by the six aforementioned parameters, the model has to simulate its interior. We define a one-dimensional spatial grid, with fixed precision, that ranges from the center of the planet to above its total radius. The internal structure of a planet is governed by its gravitational acceleration $g$, pressure $P$, temperature $T$, and density $\rho$ inside the body. The four quantities are computed for every point of the grid, by solving the canonical equations of internal structure for solid bodies. The gravitational acceleration is computed from the Gauss theorem:

\begin{equation}
	\dfrac{d g}{d r} = 4 \pi G \rho - \dfrac{2 G m}{r^3}
	\label{eq:3}
\end{equation}

\noindent with $m$ the mass at a given radius $r$, and $G$ the gravitational constant. Thus, at a radius $r \in ]R_i,R_{i+1}]$ of the spatial grid:

\begin{equation}
	g(r) = \frac{4 \pi G}{r^2} \int_{R_i}^{r} r^2 \rho \ dr + g(R_i) \left( \frac{R_i}{r} \right)^2
	\label{eq:4}
\end{equation}

\noindent where $R_i$ is the lower radius of a layer $i \in \llbracket 1,5 \rrbracket$ (plus $R_6 = R_{\rm P}$ the planet total radius), i.e. $R_1 = 0$. The pressure $P$ is then computed from the planet surface assuming the hydrostatic equilibrium:

\begin{equation}
	\dfrac{d P}{d r} = - \rho g \qquad \Longrightarrow \qquad P(r) = P(R_{i+1}) + \int_{r}^{R_{i+1}} \rho g \ dr
	\label{eq:5}
\end{equation}

\noindent as well as the temperature $T$, if we consider an adiabatic profile:

\begin{equation}
	\frac{dT}{dP} = \frac{\gamma T}{\rho \Phi} \qquad \Longrightarrow \qquad \dfrac{d T}{d r} = - g \dfrac{\gamma T}{\Phi} \qquad \Longrightarrow \qquad T(r) = T(R_{i+1}) \ exp \left[ \int_{r}^{R_{i+1}} \frac{\gamma g}{\Phi} \ dr \right]
	\label{eq:6}
\end{equation}

\noindent via the use of the Adams-Williamson equation:

\begin{equation}
	\frac{d\rho}{dr} = - \frac{\rho g}{\Phi}
	\label{eq:7}
\end{equation}

\noindent with $\gamma$ and $\Phi$ the Gr\"uneisen and seismic parameters, respectively:

\begin{equation}
	\begin{dcases}
		\gamma = \gamma_0 \left( \frac{\rho_0}{\rho} \right) ^q		\\
		\Phi = \frac{K_S}{\rho} = \frac{dP}{d \rho}
	\end{dcases}
	\label{eq:8}
\end{equation}

\noindent and $\gamma_0$ the reference Gr\"uneisen parameter, $\rho_0$ the density at ambient conditions (kg/m\textsuperscript{3}), $q$ the adiabatic power exponent, and $K_S$ the adiabatic bulk modulus (GPa). Following the approach of \cite{Sotin07}, the temperature profile is adapted to mimic the presence of thermal boundary layers at the top and bottom of each layer animated by convection, by fixing temperature drops between the different layers, whose values are taken from Earth models.

Finally, the mass of the planet verifies:

\begin{equation}
	\dfrac{d m}{d r} = 4 \pi r^2 \rho \qquad \Longrightarrow \qquad M_P = \sum_{i=1}^{5} \int_{R_i}^{R_{i+1}} r^2 \rho \ dr
	\label{eq:9}
\end{equation}

The computational scheme works as follows: the six input parameters $M_{\rm P}$, $x_{\rm core}$ and $x_{\rm water}$, $f_{\rm alloy}$, $\left( \frac{\rm Mg}{\rm Si} \right) _{\rm P}$, and $\rm Mg\#$ (see Section~\ref{ssec:2.1}) are given, but only the latter three are used at first, since they fix the distribution of materials in the different layers. The model starts with a planet composed of the five layers (core to liquid water) with lower radii $R_i$ fixed arbitrarily, and a homogeneous density fixed to the density of the corresponding material at ambient conditions $\rho_0(i)$. Then are computed the profiles of $g$, $P$, and $T$ inside the planet using Equations~\ref{eq:3}--\ref{eq:8}, followed by the profile of $\rho$ computed using the corresponding equation of state of the material (see Section~\ref{ssec:2.2}). $g$ and $\rho$ are computed from the center of the planet with an increasing radius, whereas the computation of $P$ and $T$ starts at the planet surface and is done in the opposite direction. This requires boundary conditions, namely: no central gravitational acceleration, surface pressure and temperature fixed to given values $P_{\rm surf}$ and $T_{\rm surf}$. These parameters allow to simulate the presence of an atmosphere, provided that its mass and height are negligible compared to $M_{\rm P}$ and $R_{\rm P}$ respectively (otherwise the gaseous atmosphere should be included in the model as a supplementary layer). This is the case of the Earth, where the atmosphere only accounts for 0.0001\% of the planet's mass.

From the profiles of these four quantities, we are then able to re-estimate the layers' lower radii $R_i$ to fit the leftover input parameters:

\begin{itemize}
	\item[-] $R_1 = 0$ by definition;
	\item[-] $R_2$ gives the size of the core, so we fix it using the CMF: 
	\begin{equation}
		\int_{0}^{R_2} \rho(x) r^2(x) \ dx = x_{\rm core} M_{\rm P}
		\label{eq:10}
	\end{equation}
	\item[-] $R_3$ is the boundary between the lower and upper mantles, which corresponds to the phase change of silicate rocks (from bridgmanite and ferro-periclase to olivine and enstatite; see Section~\ref{ssec:2.1}). Following the work of \cite{Sotin07}, this phase change is well described by a Simon equation \citep{Irifune87}, giving:
	\begin{equation}
		T(R_3) = T_0 + \frac{P(R_3) - P_0}{a}
		\qquad
		\text{with}
		\qquad
		\begin{cases}
			T_0 = 800 \ \text{K}										\\
			P_0 = 25 \cdot 10^9 \ \text{Pa}								\\
			a = - 0.0017 \cdot 10^9 \ \text{Pa K\textsuperscript{-1}}
		\end{cases}
		\label{eq:11}
	\end{equation}
	\item[-] as for the boundary between core and mantle, $R_4$ is the limit between (upper) mantle and hydrosphere, thus we compute it using the mass fraction of the mantle (from the CMF and WMF):
	\begin{equation}
		\int_{R_2}^{R_4} \rho(x) r^2(x) \ dx = (1 - x_{\rm core} - x_{\rm water}) M_{\rm P}
		\label{eq:12}
	\end{equation}
	\item[-] $R_5$ is then the radius at which the phase transition of ice VII to liquid water occurs, i.e. (see Section~\ref{ssec:2.4}; \citealp{Frank04}):
	\begin{equation}
		P(R_5) = P_0 + a \left[ \left( \frac{T(R_5)}{T_0} \right) ^c - 1 \right]
		\qquad
		\text{with}
		\qquad
		\begin{cases}
			T_0 = 355 \ \text{K}					\\
			P_0 = 2.17 \cdot 10^9 \ \text{Pa}		\\
			a = 0.764 \cdot 10^9 \ \text{Pa}		\\
			c = 4.32
		\end{cases}
		\label{eq:13}
	\end{equation}
	\item[-] finally, $R_6 = R_{\rm P}$ the total radius of the planet, is fixed by the mass of the hydrosphere:
	\begin{equation}
		\int_{R_4}^{R_6} \rho(x) r^2(x) \ dx = x_{\rm water} M_{\rm P}
		\label{eq:14}
	\end{equation}
\end{itemize}

The latter two steps (computation of $g$, $P$, $T$, and $\rho$, and estimation of the $R_i$) are then repeated in an iterative scheme, until convergence is reached. Convergence is achieved when the changes of the $R_i$ and the profiles of $g$, $P$, $T$, and $\rho$ from one iteration to the other are lower than a fixed precision.

\begin{figure}
\begin{center}
	\includegraphics[width=7.5cm]{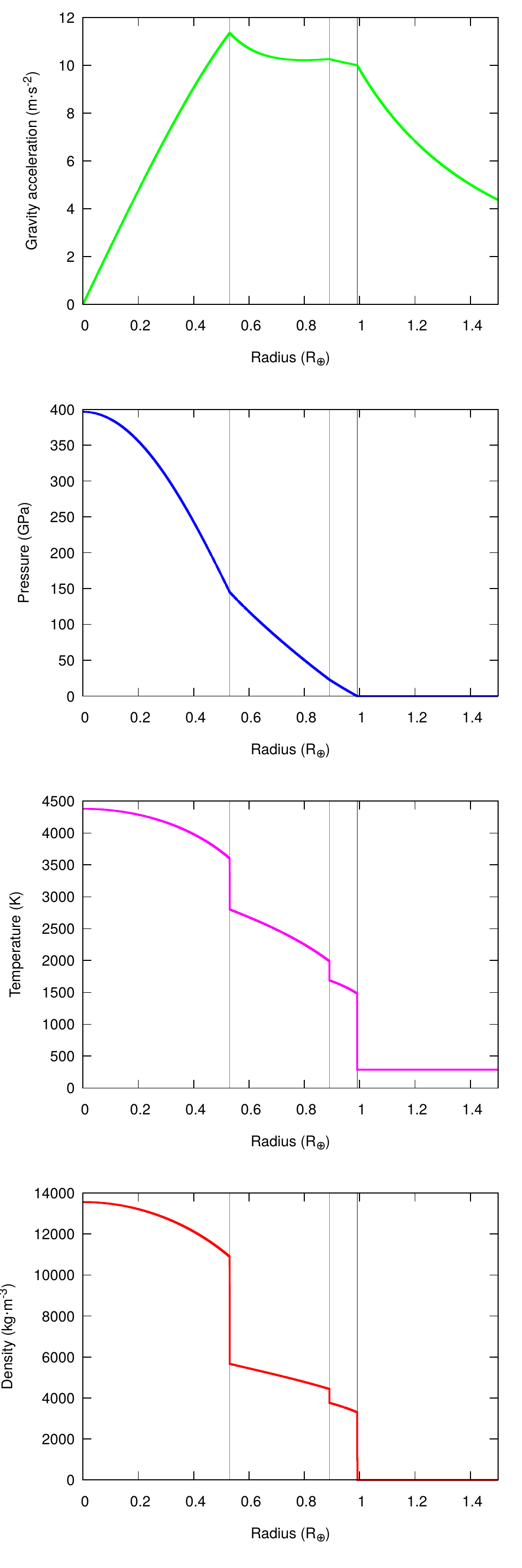}
	\caption{From top to bottom: gravity acceleration $g$, pressure $P$, temperature $T$, and density $\rho$ profiles computed inside a planet of $1\mearth$ with an Earth-like composition, once the model has reached convergence. Vertical lines show the boundaries between the different layers that compose the planet.}
	\label{fig:9}
\end{center}
\end{figure}

Once the iterative process has stopped, the model is supposed to verify all input parameters of the planet, and can provide a planet total radius $R_{\rm P}$ that respects the thermodynamic and elastic properties of the materials composing the planet. This computed radius can then be compared to the measured radius, if there is one known. We also have access to the interior profiles of $g$, $P$, $T$, and $\rho$, as represented on Figure~\ref{fig:9} in the case of the Earth again. The transitions between the different layers are easily noticed through the discontinuities of the curves.

As shown on Figure~\ref{fig:9}, simulating a 1~$\mearth$ planet with the Earth parameters (Table~\ref{tab:1}) produces a radius $R_{\rm P} = 0.992$~$\rearth$, i.e. $\sim$60~km less than the actual Earth. This less than 1\% error is comparable to those obtained by other models \citep{Sotin07,Valencia06}, since none of these models considers the Earth's crust (a 10--50~km thick layer of low density), and also because several chemical elements are not incorporated into the modeling of the core and mantles (as Ca, Al, and Ni) for sake of simplicity. Here, to obtain a planet as big as the Earth (6371~km, with an error of 0.001\%), the CMF has to be lowered to 0.286 (the value actually inferred for Venus; \citealp{Stacey05}), if all other parameters remain fixed to Earth values.

\subsection{Planetary Fe/Si ratio}

From our model, it is possible to compute output compositional parameters of the simulated planet, that were not in the set of input parameters. For instance, the Fe/Si mole ratio of the planet can be calculated using the following equation:

\begin{equation}
	\left( \frac{\rm Fe}{\rm Si} \right) _{\rm P} = \frac{\sum_{i=1}^{5} n_i (\rm Fe)_i}{\sum_{i=1}^{5} n_i (\rm Si)_i}
	\label{eq:15}
\end{equation}

\noindent with $n_i = \frac{M_i}{M_{{\rm mol},i}}$, where $M_i$ is the mass of the layer $i$, and $M_{{\rm mol},i}$ is the mean molecular mass of the material composing this layer. Here $(\rm X)_i$ is the mole fraction of the element X in layer $i$. As for the planet radius, the computed Fe/Si ratio can be compared to a measured value, or provide an estimation when the latter is unknown.

However, it is interesting to note that in our case, the Fe/Si ratio of the simulated planet can be derived analytically:

\begin{equation}
	\left( \frac{\rm Fe}{\rm Si} \right) _{\rm P} = \frac{1}{\rm Mg\#} \left( \frac{\rm Mg}{\rm Si} \right) _{\rm P} \left[ 1 - \rm Mg\# + \frac{\frac{M_1}{M_{{\rm mol},1}}}{\frac{M_2}{M_{{\rm mol},2}} + 2 \frac{M_3}{M_{{\rm mol},3}}} \right]
	\label{eq:16}
\end{equation}

Yet, there is a simple relation between the molecular masses of the mantles, namely $M_{{\rm mol},3} = 2 M_{{\rm mol},2}$. This comes from our assumptions that the mantle is chemically homogeneous (see Equations~\ref{eq:1}--\ref{eq:2}), but also from the molecules present in these layers (bridgmanite, olivine, ferro-periclase, and enstatite), whose molecular masses compensate each other. Eventually, we obtain:

\begin{equation}
	\left( \frac{\rm Fe}{\rm Si} \right) _{\rm P} = \frac{1}{\rm Mg\#} \left( \frac{\rm Mg}{\rm Si} \right) _{\rm P} \left[ 1 - \rm Mg\# + \frac{M_{{\rm mol},2}}{M_{{\rm mol},1}} \frac{x_{\rm core}}{x_{\rm mantle}} \right]
	\label{eq:17}
\end{equation}

Interestingly, the Fe/Si mole ratio of a planet does not depend on the mass of the body, only on its compositional parameters. In particular, if the parameters $f_{\rm alloy}$, $\left( \frac{\rm Mg}{\rm Si} \right) _{\rm P}$, and $\rm Mg\#$ are fixed, the Fe/Si ratio only depends on the CMF and WMF of the planet, and can thus be represented in the ternary diagram ``core-mantle-water'' (see Section~\ref{ssec:2.6}).

\subsection{Equations of state}

The thermodynamic and elastic parameters used in the EOS hereafter detailed, that describe the behavior of each material composing the planet, are taken from \cite{Sotin07} and \cite{Sotin10}.

\subsubsection{Third-order Birch-Murnaghan (BM3)}

\begin{equation}
	P(\rho,T) = \frac{3}{2} K_{T,0} \left[ \left( \frac{\rho}{\rho_{T,0}} \right) ^{\frac{7}{3}} - \left( \frac{\rho}{\rho_{T,0}} \right) ^{\frac{5}{3}} \right] \left\lbrace 1 - \frac{3}{4} (4 - K'_{T,0}) \left[ \left( \frac{\rho}{\rho_{T,0}} \right) ^{\frac{2}{3}} - 1 \right] \right\rbrace
	\label{eq:19}
\end{equation}

\begin{equation*}
	\text{with}
	\qquad
	\begin{cases}
		K_{T,0} = K_0 + a_P (T - T_0)																	\\
		K'_{T,0} = K'_0																					\\
		\rho_{T,0} = \rho_0 \ \exp \left( {\stretchint{5ex}}_{\bs T_0}^{T} \alpha(t) \ dt \right)		\\
		\alpha(T) = a_T + b_T T - c_T T^{-2}
	\end{cases}
\end{equation*}

\noindent where $T_0$, $\rho_0$, $K_0$, $K'_0$, $a_P$, and $\left\lbrace a_T, b_T, c_T \right\rbrace$ are the reference temperature, density, bulk modulus, pressure and temperature derivatives of the bulk modulus, and thermal expansion coefficients, respectively. 

\subsubsection{Mie-Gr\"uneisen-Debye (MGD)}

\begin{equation}
	P(\rho,T) = P(\rho,T_0) + \Delta P_{\text{th}}
	\label{eq:20}
\end{equation}

\begin{equation*}
	\text{with}
	\qquad
	\begin{dcases}
		P(\rho,T_0) = \frac{3}{2} K_0 \left[ \left( \frac{\rho}{\rho_0} \right) ^{\frac{7}{3}} - \left( \frac{\rho}{\rho_0} \right) ^{\frac{5}{3}} \right] \left\lbrace 1 - \frac{3}{4} (4 - K'_0) \left[ \left( \frac{\rho}{\rho_0} \right) ^{\frac{2}{3}} - 1 \right] \right\rbrace							\\
		\Delta P_{\text{th}} = 9 \frac{\gamma n R}{V \theta^3} \left[ T^4 {\stretchint{5ex}}_{\bs 0}^{\frac{\theta}{T}} \frac{t^3}{\mathrm{e}^t - 1} \ dt - T_0^4 {\stretchint{5ex}}_{\bs 0}^{\frac{\theta}{T_0}} \frac{t^3}{\mathrm{e}^t - 1} \ dt \right]
	\end{dcases}
\end{equation*}

\noindent with $\theta = \theta_0 \left( \frac{\rho}{\rho_0} \right) ^{\gamma}$ and $\gamma = \gamma_0 \left( \frac{\rho}{\rho_0} \right) ^{-q}$, $\theta_0$, $\gamma_0$, $q$, and $n$ being the reference Debye temperature and Gr\"uneisen parameters, scaling exponent, and number of atoms per chemical formula, respectively.

\subsubsection{Vinet}

\begin{equation}
	P(\rho,T_0) = 3 K_0 \left[ \left( \frac{\rho}{\rho_0} \right) ^{\frac{2}{3}} - \left( \frac{\rho}{\rho_0} \right) ^{\frac{1}{3}} \right] \exp \left\lbrace \frac{3}{2} (K'_0 - 1) \left[ 1 - \left( \frac{\rho}{\rho_0} \right) ^{- \frac{1}{3}} \right] \right\rbrace
	\label{eq:21}
\end{equation}

\noindent with addition of the thermal pressure $\Delta P$, as for the MGD formulation.

\end{document}